\documentclass[aps,prl,twocolumn,showpacs,groupeaddress]{revtex4}
\usepackage{graphicx}
\usepackage{natbib}

\newcommand{\ls}{\mathrel{\raise1.16pt\hbox{$<$}\kern-7.0pt 
\lower3.06pt\hbox{{$\scriptstyle \sim$}}}}         
\newcommand{\gs}{\mathrel{\raise1.16pt\hbox{$>$}\kern-7.0pt 
\lower3.06pt\hbox{{$\scriptstyle \sim$}}}}         

\long\def\comment#1{}

\def\fun#1#2{\lower3.6pt\vbox{\baselineskip0pt\lineskip.9pt
  \ialign{$\mathsurround=0pt#1\hfil##\hfil$\crcr#2\crcr\sim\crcr}}}

\def\ba{\begin{eqnarray}}
\def\ea{\end{eqnarray}}
\def\be{\begin{equation}}
\def\ee{\end{equation}}

\begin{document}

\title{Implications for Primordial Non-Gaussianity ($f_{NL}$) from weak lensing masses of high-z galaxy clusters.}

\author{Raul Jimenez$^{1,2}$, and Licia Verde$^{1,2}$}
\email[Email: ]{raulj, lverde@astro.princeton.edu}
\affiliation{$^{1}$ICREA \& Institute of Sciences of the Cosmos (ICC), University of Barcelona, Barcelona 08028, Spain. \\
$^{2}$Theory Group, Physics Department, CERN, CH-1211 Geneva 23, Switzerland. }

\date{\today}

\begin{abstract}
The recent weak lensing measurement of the dark matter mass of the high--redshift galaxy cluster XMMUJ2235.3-2557 of $(8.5 \pm 1.7) \times 10^{14}$ $M_{\odot}$ at $z=1.4$, indicates that, if the cluster is assumed to be the result of the collapse of dark matter in a primordial gaussian field in the standard  LCDM model, then its abundance should be $ < 2 \times 10^{-3}$ clusters in the observed area.
Here we investigate how to boost the probability of XMMUJ2235.3-2557 in particular  resorting to deviations from Gaussian initial conditions. We show that this abundance can be boosted by factors $> 3-10$ if the non-Gaussianity parameter  $f^{local}_{NL}$ is in the range $150-200$. This value is comparable to the limit for $f_{NL}$ obtained  by current constraints from the CMB.  We conclude that mass determination of high-redshift, massive clusters  can offer a complementary probe of primordial non-gaussianity.
\end{abstract}

\pacs{cosmology}

\maketitle

{\it Introduction.---}
It has been recognized for  almost  a decade that the abundance of the most massive and/or high-redshift collapsed objects could be used to constraint the nature of the primordial fluctuation field \cite{MVJ00,RGS00,Willick00,VJKM01}. The subject has recently received renewed attention \cite{LoVerde08,voids09,Grossi,Pillepich,desjaques} possibly sparked by a claimed detection of deviations from Gaussianity on CMB maps \cite{wandelt08}. Depending on the sign of the non-Gaussian perturbation, the abundance of rare objects will be  enhanced or depleted. In \cite{MVJ00} we developed the necessary theoretical tools to interpret any enhancement (depletion) in the abundance of rare-peak objects over the gaussian initial conditions  case. Working with ratios of non-Gausian  over the Gaussian case makes the theoretical predictions more robust. Later on, Ref.~\cite{LoVerde08} generalised the procedure to more modern mass-functions and type of non-gaussianity including scale dependence. The validity of the analytical formulas developed in \cite{MVJ00} has been recently confirmed by detailed N-body numerical simulations with non-gaussian initial conditions \cite{Grossi}. These authors have shown  that the analytical findings in \cite{MVJ00}  provide an excellent fit to the non-Gaussian mass function found in N-body simulations with a simple  ``calibration" procedure.

Ref.~\cite{Jee09} have  recently reported  a weak-lensing analysis of the $z = 1.4$ galaxy cluster XMMU J2235.3-2557 based in HST (ACS) images. Assuming a NFW \cite{nfw} dark matter profile for the cluster, they estimate a projected mass within $1$ Mpc of $(8.5 \pm 1.7) \times 10^{14}$ M$_{\odot}$. Adopting a LCDM cosmology  with cosmological parameters given by WMAP 5 yr data (\cite{komatsu09}) and assuming Gaussian initial conditions they estimate that in the surveyed 11 sq. deg. there should be $0.005$ clusters above that mass. Therefore the observed cluster is unlikely a the $3\sigma$ level. In this Letter we explore what level of non-Gaussianity is required to boost this abundance by a factor $\sim 10$ and how this relates to the available constraints obtained from the CMB. We show that with $f^{local}_{NL}$ in the range $150-200$ it is possible  significantly enhance  the  abundance  expected for such a massive cluster. This value of $f_{NL}$ is comparable with current limits from the CMB \cite{komatsu09} , ~\cite{wandelt08}. \\

{\it High Redshift and/or Massive Objects.---}
While there are in principle infinite types of possible deviations from Gaussianity, it is common to parameterize these deviations in terms of the  dimensionless parameter $f_{NL}$
(e.g., \cite{CB87,VWHK00,MVJ00,komatsu01}).  
\begin{equation}
     \Phi=\phi+f^{local}_{ NL}(\phi^2-\langle\phi^2\rangle).
\label{secondngmap}
\end{equation}
 where $\Phi$ denotes the {\it primordial}
Bardeen potential \footnote{Which, on sub-Hubble spaces  reduces to the usual Newtonian potential with a negative sign.} and $\phi$ denotes a Gaussian random field. With this convention a 
positive value of $f^{local}_{NL}$ will yield to a positive skewness in the density field  and an enhancement in the number of rare, collapsed objects.

Although not fully general, this model (called {\it local}-type)  may be considered as
the lowest-order terms in Taylor expansions of more general
fields. Local non-gaussianity arises in standard slow roll inflation (although in this case $f^{local}_{NL}$ is unnmeasurably small), and in multi-field models  (e.g., \cite{Luo94,FRS93,GLMM94, Fanbardeen92}).
For other types of non-gaussianity (as we will see below) an ``effective" $f_{NL}$ can be defined and related to this model. 

The abundance of  {\it rare events} (high-redshift and/or
massive objects) is determined by the form of the high-density tail of the
primordial density distribution function.  A probability distribution function
(PDF) that produces a larger number of $>3\sigma$ peaks than a Gaussian
distribution will lead to a larger abundance of rare  events.
Since small deviations from Gaussianity have deep impact on
those statistics that probe the tail of the distribution (e.g. \cite{Fry86,MVJ00}), rare events should be powerful probes of primordial non Gaussianity.

The non-Gaussianity parameter $f_{NL}$ is effectively a ``tail
enhancement'' parameter (c.f., \cite{MVJ00}).

As shown in \cite{VJKM01,LoVerde08,Grossi}  when using an analytical approach to compute  the mass function  a robust quantity to use is the fractional non-gaussian correction to the Gaussian mass function ${\cal R}_{NG} (M,z)$. This quantity was calibrated on non-gaussian N-body simulations in \cite{Grossi}. For our purpose here we want to compute a closely related quantity:  the {\it non-gaussianity enhancement} i.e.  ratio of the non-gaussian  to gaussian abundance of halos above a mass threshold \cite{VJKM01}. As the mass function is exponentially steep  for rare events here we can safely make the identification of the non-gaussianity enhancement with ${\cal R}_{NG}$.

To understand the effect of non-gaussianity on halo abundance let us recall that  
 to first order  the {\it non-gaussianity enhancement} is given by \cite{LoVerde08,Grossi}:
 \begin{equation}
 {\cal R}_{NG} (M,z)\sim 1+S_{3,M}\frac{\delta'_c(z)^3}{6\sigma_M^2}
 \end{equation}
where $S_{3,M}$ denotes the skewness of the density field linearly extrapolated at $z=0$ and smoothed on a scale $R$ corresponding to the comoving Lagrangian radius of the halo of mass $M$, $\sigma_M$ denotes the {\it rms} if the --linearly extrapolated at $z=0$-- density field  also smoothed on the same scale $R$; $\delta'_c(z_f)=\sqrt{q}\delta_c(z_f)$  and $\delta_c(z_f)$ denotes critical collapse density at the formation redshift of the cluster $z_f$. Note that $\delta_c(z)=\Delta_cD(z=0)/D(z)$ with $D(z)$ denoting the linear growth factor and $\Delta_c$ is a quantity slightly dependent on redshift and on cosmology, which  only for an Einstein-de-Sitter Universe is  constant $\Delta_c =1.68$. The constant $q\simeq 0.75$ (which we will call ``barrier factor") can be physically understood as the effect of non-spherical collapse \cite{LeeShandarin98,ShethMoTormen99} lowering the critical collapse threshold of a diffusing barrier \cite{riotto} see also \cite{Robertson08}, and has been   calibrated on N-body simulations in Ref.~\cite{Grossi}. The full expression for ${\cal R}_{NG}$ is (cf Eqs. 6 and 7 in Ref.~\cite{Grossi}): 

\begin{eqnarray}
\label{eq:ratioMVJellips}
&&{\cal R}_{NG}(M,z)=\exp\left[(\delta'_{c})^3
\frac{S_{3,M}}{6 \sigma_M^2}\right] \times \\
& &\!\!\!\! \left| \frac{1}{6}
\frac{\delta_{ec}^2}{\sqrt{1-\delta'_{c}S_{3,M}/3}} 
\frac{dS_{3,M}}{d\ln \sigma_{M}} \right. 
+\left. \frac{\delta'_{c}\sqrt{1-\delta'_{c} S_{3,M}/3}}{\delta'_{c}}\right|  \,.
\nonumber 
\end{eqnarray}

Let us re-iterate that in principle the enhancement factor should be computed by integrating the mass function $n(M,z,f_{NL})$ between the minimum and the maximum mass and for redshifts above the observed one \cite{VJKM01}:
\begin{equation}
\widehat{\cal R}_{NG}=\frac{\int n(M,z,f_{NL}) dM dz}{\int n(M,z,f_{NL}\equiv 0)dM dz}
\end{equation}
but since the mass function, in the regime we are interested in, is exponentially steep, we can identify $\widehat{\cal R}_{NG}={\cal R}_{NG}$.

\begin{figure}[ht!]
\includegraphics[width=\columnwidth,angle=0]{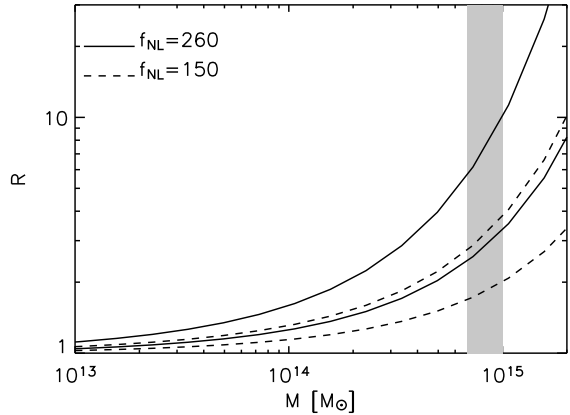}
\caption{Enhancement factor ${\cal R}_{NG}$ of the number of rare objects for different values of the dark matter mass of the galaxy cluster. The lines correspond to different values of $f_{NL}$. The upper lines are for a collapse redshift of $z_f=2$ and the lower lines for $z_f=1.4$. The shaded area is the range for the weak lensing mass estimate of the clusters XMMUJ2235.3-2557. Note that for the quoted values of $f^{local}_{NL}$ it is possible to obtain enhancements of order $10$ in the cluster number abundance. This enhancement brings the expected abundance of such massive clusters in better agreement with the observations. Note that for masses above the estimated central value ($8.5 \times 10^{14}$ M$_{\odot}$) one expects to find {\it zero} such objects in the {\it whole} sky (one expects 7 objects in the whole sky at the lowest value of the mass estimate) which emphasizes the need of an enhancement as the one provided by primordial non-gaussianity studied here. \label{f1}}
\end{figure}

Small deviations from Gaussian initial conditions will lead to a non-zero skewness and in particular for local non Gaussianity  $S_{3,M}= f^{local}_{NL} S_{3,M}^{1}$ where $S_{3,M}^{1}$ denotes the skewness produced by $f^{local}_{NL}=1$. Since non-Gaussianity comes in  the expression for 
${\cal R}_{NG} $ only through the skewness, the same expression can be used for other types of non-Gaussianity such as the {\it equilateral} type (see e.g. Ref.~\cite{LoVerde08,voids09} for example of applications). For example, at the scales of interest $R=13 Mpc/h$,  $S_{3,R}^{1,local} = 3.4 S_{3,R}^{1,equil}$ thus when working on these scales  to obtain the same non-Gaussian enhancement as  a local  model, an equilateral model needs  a higher effective $f_{NL}$:  we can make the identification $f_{NL}^{equil}= 3.4 f_{NL}^{local}$.

Here we will use the full  \cite{MVJ00} expression, corrected for the ``barrier factor", for the non-gaussian mass function to compute the  {\it non-gaussianity enhancement}. Note that  the estimated  mass and redshift of XMMUJ2235.3-2557, places it just outside the range where the mass function expressions of \cite{MVJ00, LoVerde08} have been directly reliably tested with non-Gaussian N-body simulations. Simulations seems to indicate that the \cite{MVJ00} expression is a better fit than \cite{LoVerde08} at high masses/redshift and large $f_{NL}$, this is also supported by theoretical considerations \cite{Grossi}. \\

{\it Results.---}
Fig.~\ref{f1} shows the enhancement factor ${\cal R}_{NG}$ as a function of the mass of the galaxy cluster for different values of $f^{local}_{NL}$ and the redshift of collapse. The shaded area shows the error band for the mass determination of XMMUJ2235.3-2557 from Ref.~\cite{Jee09} and  the different lines have been computed using the \cite{MVJ00} mass function, with the ``barrier factor" correction. Ref.~\cite{Grossi} show that  it  fits very well the N-body numerical simulations for the case of rare peaks, which is the one we are concerned with. The solid lines correspond to $f_{NL}=260$, the lower one is for a cluster collapse redshift of $z_f=1.4$ (i.e. assuming that  the cluster forms at the observed redshift) and the upper one for $z_f=2$. The two dashed lines also depict the mentioned collapse redshifts but for $f_{NL} = 150$. We see that the galaxy cluster abundance  can be enhanced by a factor up to $10$. In the mass range of interest, the same enhancement factor can be obtained for an equilateral-type non-gaussianity for $f_{NL}^{equil} =884$ and $510$ respectively.

We should bear in mind that XMMUJ2235.3-2557 is an extremely rare object, sampling the tail of the mass function which may not be well known  and may be strongly affected by cosmology. 
Using the \cite{ShethTormen99} mass function we estimate that in the WMAP5  LCDM model \footnote{The cosmological parameter which uncertainty  affect most the cluster abundance is $\sigma_8$, the value adopted is $\sigma_8=0.77$. } one should find $7$ galaxy clusters in the whole sky with mass  greater or equal than the lower mass estimate of XMMUJ2235.3-2557  $M  = 5\times 10^{14} M_{\odot}$  and $z > 1.4$ corresponding to a probability of $0.002$ for the $11$ deg$^2$ of the survey. This should be compared with the reported number of $0.005$ obtained  by \cite{Jee09} for a different cosmology and different mass function. Thus the effects of  cosmology and uncertainty in the mass function can account for  a factor $\sim 2$ uncertainty in the predicted halo abundance.

Note that in all our calculations we have used a conservative lower limit for the mass of the cluster. If instead we use the central or upper value for the mass, using the WMAP5 cosmology and the \cite{ShethTormen99} mass function we expect to find {\it zero} such clusters in the whole sky, which will make our conclusions even stronger.    
 
The survey area used in Ref.~\cite{Jee09} is  $11$ deg$^2$, but the XMM serendipitus survey in 2006  covered $168$ deg$^2$ and today covers  $\sim 400$ deg$^2$. Below we report  the Ref.~\cite{Jee09} numbers and in parenthesis the numbers we obtain. 
The probability of finding  XMMUJ2235.3-2557 is thus  $0.005$ ($0.002$) if using  $11$ deg$^2$; to avoid as much as possible biases due to {\it a posteriori} statistics one could use  $168$ deg$^2$ obtaining a probability of  $0.07$ ($0.03$), or, as a limiting case,  even $0.17$ ($0.07$) if using  $400$ deg$^2$. Note that it is likely that there are more clusters as massive in the survey area \cite{Stern} and therefore these  numbers are  conservative.   If we use from Fig.~\ref{f1} the factor $3$ to$10$ enhancement, we find that it would help boost the probability to   $\sim 1$  in the surveyed areas.

The latest WMAP compilation \cite{komatsu09} reports  $-9 < f^{local}_{NL}< 111$ and  $-151 < f^{equil}_{NL} < 253$ at 95\% confidence, \cite{wandelt08} reports $27<f_{NL}^{local}<147$. The CMB however probes much larger scales ($R>120Mpc/h$)  than those probed by  clusters such as XMMUJ2235.3-2557 $R\sim 13$Mpc/$h$: a scale-dependent $f_{NL}$ with $k\sim-0.3$ can yield an effective $f_{NL}$ on  dependence XMMUJ2235.3-2557 scales that is larger than the CMB one  by a factor of $3$. 

The $f_{NL}^{local}$ values needed to accomodate the observed cluster at $z=1.4$ is in the range $150$ to $260$. This is comparable to the limits quoted by Ref.~\cite{komatsu09} and \cite{wandelt08}.\\

{\it Conclusions.---}
Accurate masses of high-redshift clusters are now becoming available through weak lensing analysis of deep images. As already discussed in previous papers \cite{MVJ00,LoVerde08}, their abundance can be used to put constraints on primordial  non-gaussianity. 
$f_{NL}^{local}$ in the range $150-260$ can boost the expected number of massive ($>5\times 10^{14} M_{\odot}$) high redshift ($z>1.4$) clusters by factors of $3$ to $10$. Such large numbers would help  make clusters like  XMMUJ2235.3-2557 much more probable. 
The scales probed by clusters are smaller than the CMB scales, and in principle non-Gaussianity may be scale-dependent,  making this a complementary approach.

 The adopted error range in the mass determination of XMMUJ2235.3-2557 is 100\%; even with such a large mass uncertainty and  considering  the pessimistic estimate  of $7$ such objects expected in the entire sky with a Poisson error of $\pm 2.6$, if the entire sky could be covered, $f^{local}_{NL}\sim 150$ could be detected at $> 4 \sigma$ level. 
 
 RJ and LV acknowledge support of MICINN grant AYA2008-03531.
LV acknowledges support of FP7-PEOPLE-2002IRG4-4-IRG\#202182. RJ is supported by a FP7-PEOPLE-IRG grant.

\end{document}